\documentclass{emulateapj}

\usepackage{graphicx}
\usepackage{epstopdf}

\begin{document}

%% one line. However, you may use \\ to force a line break if you
%% desire.

\title{Two New Halo Debris Streams in the Sloan Digital Sky Survey}

\author{C. J. Grillmair}
\affil{Spitzer Science Center, 1200 E. California Blvd., Pasadena,  CA 91125}
\email{carl@ipac.caltech.edu}

\begin{abstract}

Using photometry from Data Release 10 of the northern footprint of the
Sloan Digital Sky Survey, we detect two new stellar streams with
lengths of between $25\arcdeg$ and $50\arcdeg$. The streams, which we
designate Hermus and Hyllus, are at distances of between 15 and 23
kpc from the Sun and pass primarily through Hercules and Corona
Borealis. Stars in the streams appear to be metal poor, with [Fe/H]
$\sim -2.3$, though we cannot rule out metallicities as high as [Fe/H]
= -1.2.  While Hermus passes within $1\arcdeg$ (in projection) of the
globular cluster NGC 6229, a roughly one magnitude difference in
distance modulus, combined with no signs of connecting with NGC 6229's
Roche lobe, argue against any physical association between the
two. Though the two streams almost certainly had different
progenitors, similarities in preliminary orbit estimates suggest that
those progenitors may themselves have been a product of a single
accretion event.

\end{abstract}



\section{Introduction}

The number of known nearby halo streams (i.e. streams that we can
trace in configuration space over an appreciable portion of their
orbits around the Galaxy) currently stands at 18 \citep{grillmair2010,
  bonaca2012, grill2013, koposov2014, martin2014}.  Since we can in principle 
measure all six phase space coordinates for the stars in such
streams, they have the potential to significantly improve our
understanding of the mass distribution in the Galaxy, particularly in
regions where we have no comparable tracers \citep{deg2014, law2009,
  newberg2010, koposov2010}.  The coldest streams (e.g. Palomar 5,
GD-1, with velocity dispersions of $ \le 5$ km/s) are particularly
interesting as they may provide a sensitive means of detecting the
presence and abundance of dark matter subhalos \citep{carlberg2009,
  yoon2011, carlberg2013}.

In this paper we reexamine a portion of the northern imaging footprint
of the Sloan Survey. Using photometrically filtered star counts we
detect two long, nearly parallel overdensities whose orientations,
morphologies, and color-magnitude distributions are consistent with
stellar debris streams at distances of 18 and 21 kpc.  We briefly
describe our analysis in Section \ref{analysis}. We characterize the
new streams in Section \ref{discussion} and we put preliminary
constraints on their orbits in Section \ref{orbit}. We make concluding
remarks in Section \ref{conclusion}.

\section{Data Analysis} \label{analysis}

Photometric data in $g, r$, and $i$ for stars with $g < 22.5$ were extracted
from the SDSS DR10 release \citep{ahn2014}. We dereddened the photometry as a function
of sky position using the DIRBE/IRAS dust maps of \citet{schleg98},
corrected using the prescription of \citet{schlafly2011}. We
constructed color-magnitude filters in {\it g-r} and {\it g-i} using
the observed main sequence and red giant branch color-magnitude locus
of the metal poor globular cluster NGC 5053. The use of matched
filters for highlighting particular stellar populations has been
described at length by \citet{rock2002} and \citet{grill2009}.

The color-magnitude distribution of field stars was sampled over most
of the survey area, with the exception of regions occupied by known
streams (see Grillmair 2010 for a list). We applied the filters
to the entire survey area, and the resulting weighted star counts were
summed by location on the sky, using pixels $0.2\arcdeg \times
0.2\arcdeg$ in size, to produce two dimensional, filtered surface
density maps. The {\it g-r} and {\it g-i} maps were then coadded to improve
the signal-to-noise ratio.

To identify features caused by calibration or completeness
discontinuities between survey scans, we chose to work in the Sloan survey
coordinates $\lambda$ and $\eta$.  Discontinuities between scans are
then confined to run primarily along the $\lambda$ coordinate and can
more easily be identified and discounted.

The surface density of the filtered field star population was modeled
by first masking the new streams, and then smoothing the 2d surface
density map with a Gaussian kernel of width $7\arcdeg.$ This smooth
background model was then subtracted from the original and the result,
smoothed with a kernel of width $1\arcdeg,$ is shown in Figure 1. Note
that this figure has been optimized for a distance of 18 kpc by
shifting the NGC 5053-based color magnitude filters faintward by 0.2
magnitudes.

\section{Discussion} \label{discussion}

The two new streams appear as long, nearly parallel, nearly
great-circle enhancements extending through the constellations
Hercules and Corona Borealis.  The streams show no correlation
with reddening maps of the region \citep{schleg98} and stand out strongly 
when blinking against surface density maps optimized for
different distances or metallicities. The streams appear
somewhat fragmented, with numerous lumps and gaps which may be products
of either epicyclic motions of constituent stars
(e.g. K{\"u}pper et al.2012) or of encounters with massive structures
in the halo and/or disk. Following Grillmair (2009), we designate the
new streams Hermus and Hyllus, after two neighboring rivers in {\it
  The Illiad}. At their southern ends, both Hermus and Hyllus appear
to run off the edge of the survey footprint. Hermus extends completely
across the footprint and terminates at its northern edge, while Hyllus
appears to peter out at $\eta \approx 22\arcdeg$.

The southernmost $10\arcdeg$ of Hermus curves in a direction opposite
to that of the northern $30\arcdeg.$ Combined with a rather indistinct
and confused region connecting the southern and northern portions,
this creates some uncertainty as to whether the two features are part
of the same stream. The changing curvature is due primarily to the
coordinate system used. However, we find that while it is easy to find
orbits that fit either the northern or southern portions of the stream
individually, attempts to fit both sections simultaneously result in
significantly poorer fits, with the southern portion of the predicted
orbit lying $\approx 3\arcdeg$ to the west of the stream.  This may be a
consequence of the fact that tidal streams do not lie precisely along
a single orbit \citep{eyre2011}, or of inadequacies in our adopted
model of the Galaxy. However, it may also be that, despite 
striking similarities in distance and orientation, the northern and
southern portions of Hermus are physically unrelated. Kinematic
information will be required before we can further test this hypothesis.

The northern $30\arcdeg$ of Hermus can be well-described in J2000 equatorial
coordinates (to within $0.2\arcdeg$) using a polynomial of the form:

\begin{eqnarray}
\label{hermus_north_trace}
\alpha = 237.680 +0.75025 \delta - 0.03802 \delta^2 \nonumber \\ + 0.0005836 \delta^3
\end{eqnarray}

\noindent The entire $50\arcdeg$ of Hermus requires a higher-order fit, and can
be described to within $0.3\arcdeg$ using:

\begin{eqnarray}
\label{hermus_trace}
\alpha = 241.571 + 1.37841 \delta - 0.148870 \delta^2 \nonumber \\ + 0.00589502
\delta^3 - 1.03927 \times 10^{-4} \delta^4 \nonumber \\ + 7.28133 \times 10^{-7}
\delta^5
\end{eqnarray}

\noindent Though Hyllus shows no appreciable curvature in Figure 1,
the transformation to J2000 equatorial coordinates requires a second
order fit to match the stream to within $0.1\arcdeg:$

\begin{equation}
\label{hyllus_trace}
\alpha = 255.8150 - 0.78364 \delta + 0.01532 \delta^2
\end{equation}

To estimate the significance of the streams above the background, we
employ the T-statistic of \citet{grill2009}, which measures the median
contrast along its length between a stream and the surrounding
field. Figure 2 compares the signals of the two streams with the field
extending $15\arcdeg$ to the west (Hermus) and $9\arcdeg$ to the east
(Hyllus). Both Hermus and Hyllus are detected at the $\approx
13\sigma$ level. The lateral full-width-at-half-maxima of Hermus and
Hyllus are $\approx 0.7\arcdeg$ and $1.2\arcdeg,$ respectively. We note
that the broad width of Hyllus appears to be influenced by two or more
strong features along its length that may or may not be related to the
stream. If we ignore these features and measure the width at less
confused portions of the stream, we find a FWHM of $\approx
0.5\arcdeg~$. At mean distances of 18 and 21 kpc (see below), these
correspond to spatial widths of $\approx 220$ and 180 pc,
respectively. These widths are considerably narrower than the $>1$ kpc
widths associated with presumed dwarf galaxy streams
\citep{maje2003,grill2006a, grill2006b, belokurov2006, belokurov2007,
  grill2009}, and we conclude that the progenitors of Hermus and
Hyllus were most likely globular clusters.  The stream widths are
broader than several known or presumed globular clusters streams
(e.g. Pal 5, GD-1) but are similar to the $\approx 170$ pc width of
the Eastern Banded Structure, which \citet{grillmair2011} suggested is
a globular cluster stream which has been subject to significant
heating over time \citep{carlberg2009}.

Hermus passes within about $1\arcdeg$ of NGC 6229, a globular cluster
visible in Figure 1 near the northern edge of the SDSS footprint. However,
Hermus appears to continue past NGC 6229 with no evidence for the
characteristic S-curve that would be expected to connect to both NGC
6229 and to a trailing stream on the opposite side of the cluster.
Moreover, the published distance of NGC 6229 is 30 kpc
\citep{harris1996}, well outside the error bound for our distance
estimate for Hermus. NGC 6229 is most pronounced in our surface
density maps when we use a filter shifted at least one magnitude fainter
than a filter that optimizes the strength of Hermus itself. We
conclude that Hermus and NGC 6229 are not physically associated.

Color-magnitude diagrams (CMDs) for Hermus and Hyllus are shown in
Figure 3. These distributions were determined by using Equations 2 and
3 to select stars within $1\arcdeg$ of the centerline of each
stream. Similar regions, $4\arcdeg$ wide on either side of the stream
and laterally offset by $5\arcdeg$, were used to sample the field star
population. Scaling the latter to the former by area, the CMDs in
Figure 3 are a subtraction of the two. While not strong, main sequence
loci are clearly visible. In both cases, most of the signal in Figure
1 relies on stars with $g > 21$. Experimentation shows that the stream
signals are maximized when using a filter constructed from the CMD
locus of NGC 5053, which has [Fe/H] = $-2.29$
\citep{harris1996}. However, examination of Figure 3 suggests that we
cannot exclude metallicities as high as [Fe/H] = $-1.2$. Spectroscopy
will be required to better constrain the metallicities in these
streams.

We estimate the distances to the streams by shifting our filter
brightward and faintward. We use only the portion of the NGC 5053
locus fainter than one magnitude below the main sequence turn-off to
avoid introducing uncertainties due to the unknown ages of the stars
in these streams.  We find the northern end of Hermus peaks at a
magnitude offset relative to NGC 5053 of $-0.2 \pm 0.2$ mag, the
central portion ($\delta = 40\arcdeg$) at $0.4 \pm 0.2$ mag, while the southern
end peaks at a relative offset of $+0.3 \pm 0.2$ mag.  Adopting a distance to
NGC 5053 of 16.9 kpc \citep{harris1996}, we find that the
northern end of the stream is at a heliocentric distance of $15 \pm 3$
kpc, the central portion at $20 \pm 3$ kpc, and the southern end is at
$19 \pm 3$ kpc.

For Hyllus, we find offsets of $+0.2 \pm 0.2$ and $+0.7 \pm 0.2$ mag
at the northern and southern ends, respectively. This yields estimated
distances of $18.5 \pm 3$ and $23 \pm 3$ kpc.

Integrating the background-subtracted, {\it unfiltered} star counts
over a width of $1\arcdeg$ and within $3\sigma$ of the shifted NGC
5053 CMD locus, we find the total number of stars with $g < 22.5$ in
Hermus to be $320 \pm 190$, while for Hyllus we find $200 \pm 150$ stars.  The
large uncertainties reflect primarily Poisson statistics
due to the relatively large number of field stars at this Galactic latitude
($30 < b < 45$). To $g = 22.5$, the average surface density of the
stream stars is therefore $\approx 9 \pm 5$ stars deg$^{-2}$ for
Hermus, and $16 \pm 12$ stars deg$^{-2}$ for Hyllus. Within each
stream are clumps with surface densities in excess of 40 stars deg$^{-2}$.

\subsection{Constraints on Orbits} \label{orbit}

Given our distance uncertainties and the lack of velocity or proper
motion information, we can place only limited constraints on the
orbit. Moreover, \citet{odenkirchen2009} and \citet{eyre2011} have
demonstrated that tidal streams do not precisely trace the orbits of
their progenitors. Nevertheless, we estimate the orbits of the stream
stars to determine whether the streams might plausibly be related to
one another.  We do this using the Galactic model of \citet{allen91}
to compute trial orbit integrations, and matching these orbit
integrations with the measured positions and distances of the streams
in a least-squares sense.  We fit to 11 (Hyllus) and 12 (Hermus)
$\alpha, \delta$ normal points chosen to lie along the centerline of
each stream, and distance estimates to the centers and each
end of the streams. For Hermus we use normal points and distances for
only the northern $30\arcdeg.$ We adopt a solar Galactocentric
distance of 8.5 kpc, and stream distances as given above.  Though we
examined the proper motions of stars along the streams
\citep{munn2004, munn2008}, the relatively large uncertainties and
severe contamination by field stars combined to prevent us from
discerning any obvious stream signals.  We have therefore left the
proper motions as free parameters in the fit.

We compute orbits over a grid of radial velocity and $\mu_\alpha$,
$\mu_\delta$ proper motions to produce a $\chi^2$ data cube. $\chi^2$
is computed using the positional and distance offsets of each trial
orbit from the normal points and distance estimates above.
Uncertainties of $0.3\arcdeg$ and 3 kpc are assigned to the
positional normal points and distances, respectively. An initial grid
is computed over the ranges $-300 < v_{rad} < +300$ km s$^{-1}$ and
$-5 < \mu_\alpha, \mu_\delta < +5$ arcsec yr$^{-1}$ to locate all 
reasonable minima. Once the deepest minimum is identified, a finer
grid with 5 km s$^{-1}$ and 0.02 arcsec yr$^{-1}$ spacing is used to
find the minimum $\chi^2$ values for radial velocity and proper
motion. Uncertainties are estimated using the marginal $\chi^2$
distributions.

For Hermus, the best-fitting prograde orbit models predict a radial
velocity of $-30 \pm 30$ km s$^{-1}$, and proper motions $\mu_\alpha
=-2.52 \pm 0.03$ arcsec yr$^{-1}$, $\mu_\delta = -2.22 \pm 0.1$ arcsec
yr$^{-1}$ at a fiducial point with coordinates [$\alpha, \delta$] =
[$247.1667\arcdeg, +44.7455\arcdeg$]. A retrograde orbit would have $v_r
= -269 \pm 30$ km s$^{-1}$, $\mu_\alpha = -1.28 \pm 0.03$ arcsec
yr$^{-1}$, and $\mu_\delta = -0.05 \pm 0.1$ arcsec yr$^{-1}$ at the
same point.  The uncertainties correspond to the 95\% confidence
interval. At the northernmost point of the stream ($\delta =
50\arcdeg$) the predicted
distance and radial velocity are 13 kpc and -15 km s$^{-1}$, while at
the southernmost point ($\delta = 5\arcdeg$) the corresponding values are 19 kpc and -92 km
s$^{-1}$. Varying the radial velocities and proper motions over all
combinations of their respective uncertainty ranges, we find an
apogalactic radius of R$_a = 17.2 \pm 1.5$ kpc, perigalactic radius
R$_p = 7.1 \pm 1.0$ kpc, orbital eccentricity $e = 0.41 \pm 0.04$, and
orbital inclination $i = 42.0\arcdeg \pm 1.0\arcdeg$.

For Hyllus the best-fitting model predicts $v_r = -10 \pm 20$ km
s$^{-1}$, $\mu_\alpha = -1.82 \pm 0.02$ arcsec yr$^{-1}$, $\mu_\delta
= -1.92 \pm 0.08$ arcsec yr$^{-1}$ at coordinates [$\alpha, \delta$ ]
= [$246.9207\arcdeg, 34.1556\arcdeg$] in the prograde case. A
retrograde orbit would have $v_r = -253 \pm 20$ km s$^{-1}$,
$\mu_\alpha = -1.43 \pm 0.02$ arcsec yr$^{-1}$, and $\mu_\delta = -0.64
\pm 0.08$ arcsec yr$^{-1}$.  At the southern end of the stream the
best-fit model predicts a distance and radial velocity of 21 kpc and
-128 km s$^{-1}$.  The corresponding orbital parameters have R$_a =
18.6 \pm 1$ kpc, R$_p = 5.4 \pm 0.5$ kpc, $e = 0.55 \pm 0.03$, and
orbital inclination $i = 39.0\arcdeg \pm 0.5\arcdeg$. 

Comparing the orbital parameters of Hermus and Hyllus, we see that the
apo- and peri-Galactic distances are within $2\sigma$ of one
another. One the other hand, the well-measured orbital inclinations
are separated by at least $3\sigma$. Given our current lack of
velocity information, it would be unwise to put much weight on these
orbit fits or their differences. The fact that we see two distinct,
well-separated streams indicates that Hermus and Hyllus originated in
two different clusters. However, we cannot rule out that the parent
clusters themselves were not part of a larger accretion event. This is
made more evident in Figure 4, where we compare the best-fit orbit
projections of the two streams. The orbital parameters for such a
larger, accreted object would presumably lie somewhere near those
of Hermus and Hyllus, though the initial orbit may have been
considerably more energetic, only evolving toward the current
configuration through dynamical friction. As dynamical friction is
proportional to $1/v^2$ (where $v$ is the velocity of the progenitor
relative to the field stars in the Galaxy) this evolution would
presumably have been considerably more rapid for a prograde
orbit. Radial velocity measurements will be necessary before we can
ascertain the direction of motion, refine the orbital parameters, and establish
the degree to which the two streams might be related.

\section{Conclusion} \label{conclusion}

Hermus and Hyllus add to a growing list of debris streams that, with
suitable follow-up, will ultimately enable detailed tomography of the
Galactic halo. Refinement of the orbits of the streams will require
radial velocity and proper motion measurements of carefully selected
stars along the length of each stream. Given the relative proximity of
the streams, and if a sufficient number of stars can be found with $V
< 21$, then the Gaia survey may help to identify individual stars in
the streams (for follow-up spectroscopy), and constrain their proper
motions.  An initial survey reveals a significant number of RR Lyrae
lying in projection along both Hermus and Hyllus. If some number of
these RR Lyrae can be physically associated with the streams, then
follow-up infrared observations with the Spitzer Space Telescope would
allow us to determine much more accurate distances to various portions
of the streams. These distance measurements will in turn reduce the
uncertainties in the space motions of the streams determined by Gaia.

\acknowledgments

We are grateful to an anonymous referee for a careful reading and many
useful comments that significantly improved the final manuscript. Funding
for the creation and distribution of the SDSS Archive has been
provided by the Alfred P. Sloan Foundation, the Participating
Institutions, the National Aeronautics and Space Administration, the
National Science Foundation, the U.S. Department of Energy, the
Japanese Monbukagakusho, and the Max Planck Society.

{\it Facilities:} \facility{Sloan}.

\clearpage

\begin{figure}
%\epsscale{1.0}
\plotone{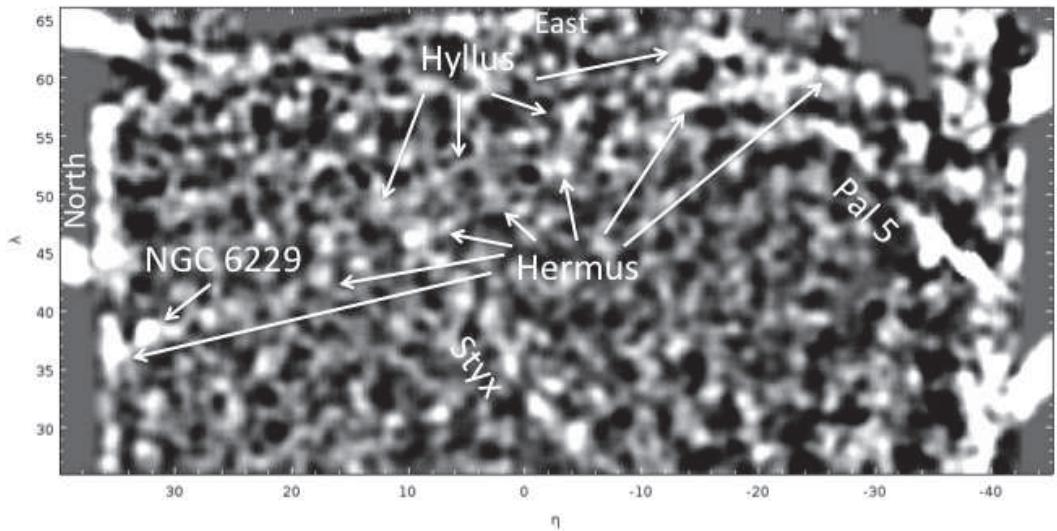}
\caption{Filtered surface density map of the eastern portion of the
  northern footprint of the Sloan imaging survey. East is upwards and
  north is to the left. The map was generated using a matched-filter
  based on the color-magnitude distribution of stars in NGC 5053,
  shifted to a distance of 18 kpc. The map has been
  background-subtracted and smoothed as described in the text. The
  Sloan survey coordinate system is used so that survey scans (and any
  discontinuities which could be mistaken for streams) run vertically
  in the figure. The stretch is linear and lighter areas indicate
  higher surface densities. Previously discovered streams
  \citep{odenkirchen2001, grill2009} are indicated.}

\end{figure}

\begin{figure}
%\epscale{1.0}
\plotone{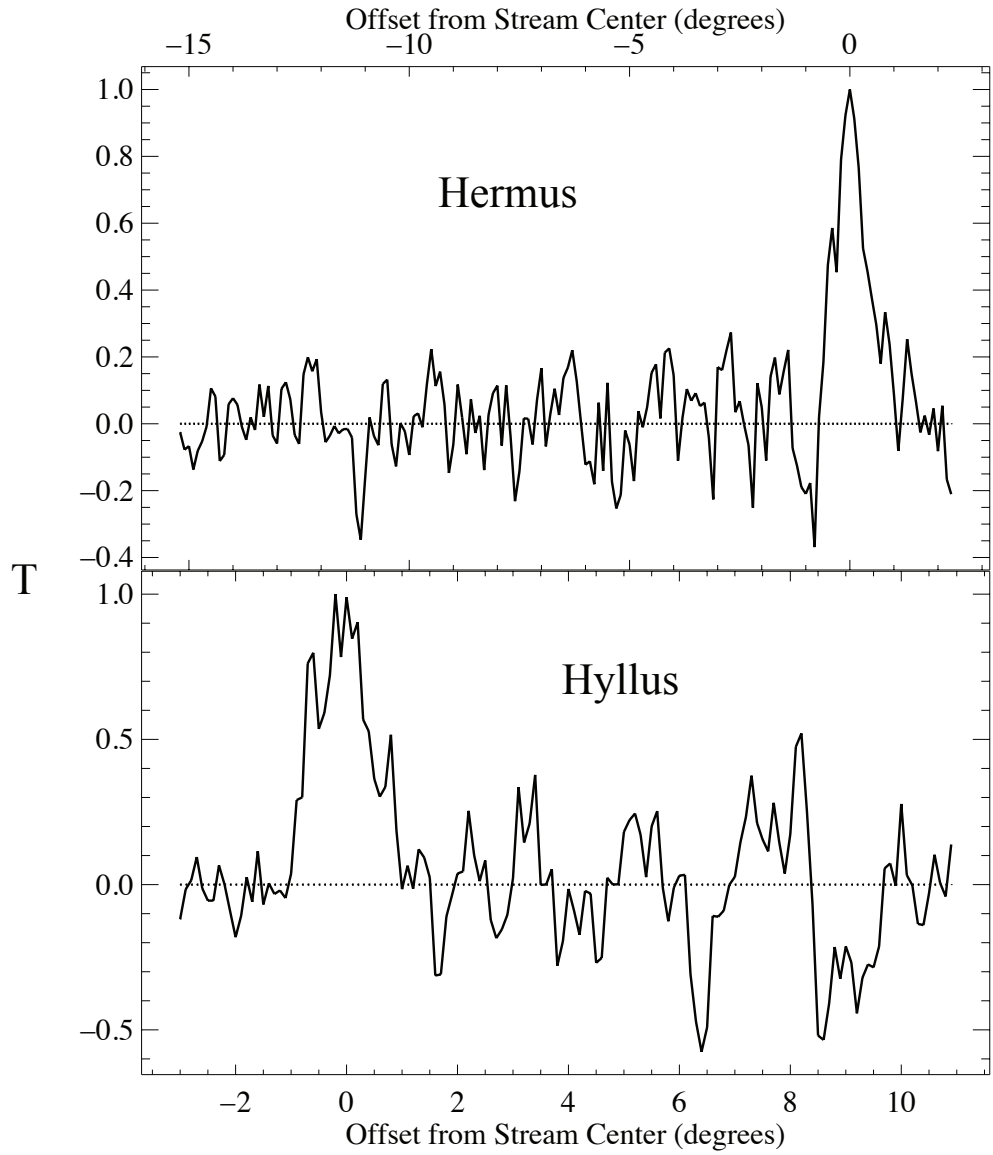}
\caption{The T-statistic \citep{grill2009}, showing the
  background-subtracted, median filtered signal over five segments,
  integrated over a width of $0.8\arcdeg,$ as a function of lateral
  offset from Hermus (top) and Hyllus (bottom).  The peak values are
  $\approx 13$ times larger than the RMS measured for the identically
  sampled regions to the west (Hermus) and east (Hyllus) of the
  streams, indicating a very low probability that the streams could be due
  to purely random fluctuations in the field. }
\end{figure}

\begin{figure}
%\epsscale{1.0}
\plotone{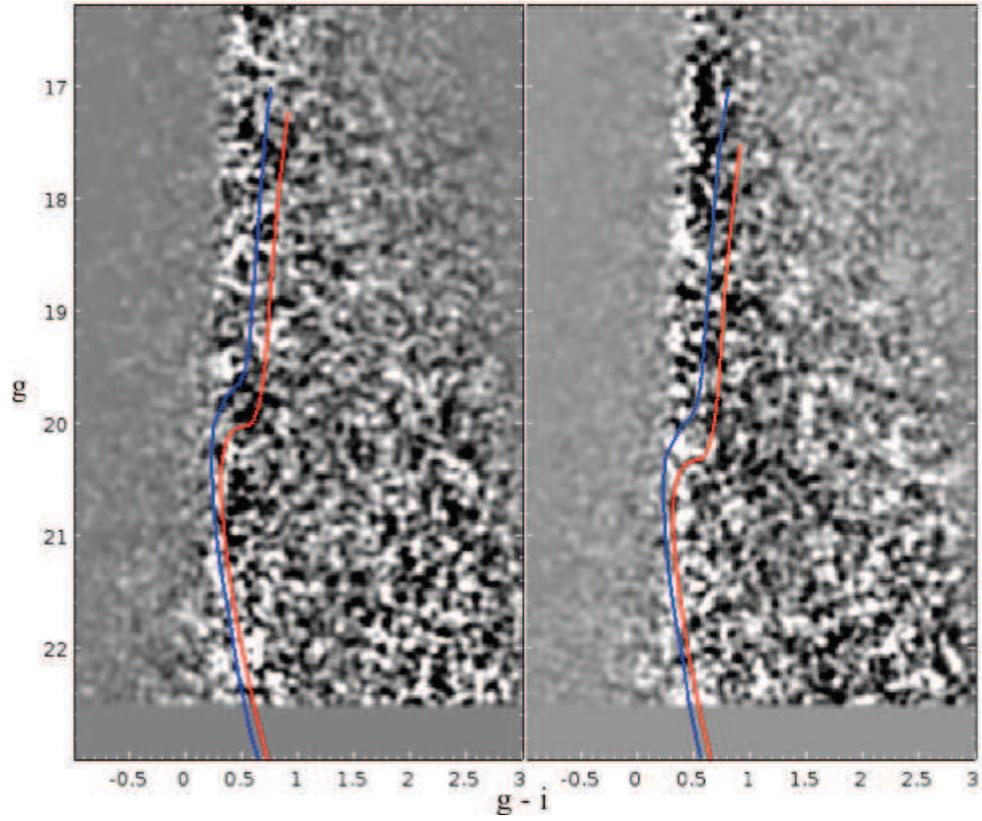}
\caption{($g, g-i$) Hess diagrams of the stars lying within $1\arcdeg$ of
  the centerlines of Hermus (left) and Hyllus (right). The blue lines
  show the shifted main sequence and red giant branch locus of the globular
  cluster NGC 5053, while the red lines show similar isochrones for
  [Fe/H] = $-1.2$. Lighter areas indicate higher surface densities.}
\end{figure}

\begin{figure}
%\epsscale{1.0}
\plotone{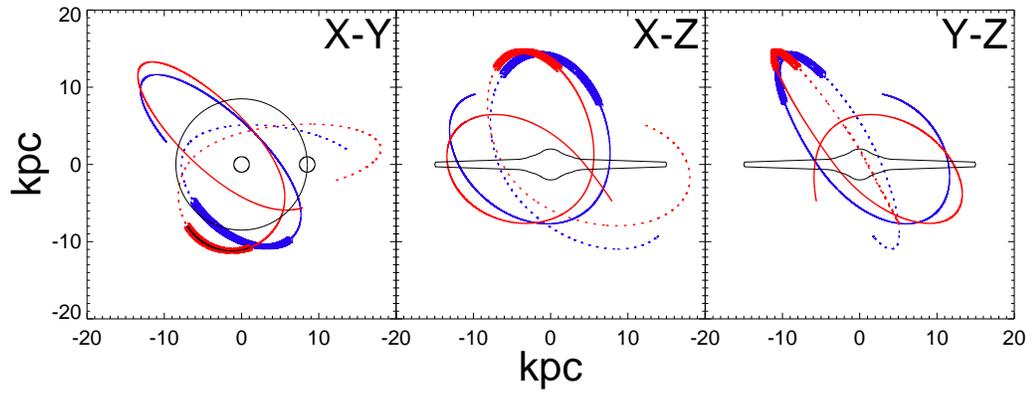}
\caption{Best-fit orbit projections for Hermus and Hyllus in X, Y, Z
  Galactic coordinates. The heavy lines show the portions of the orbits
  visible in Figure 1. The thin solid curves show the best-fit
  orbits integrated backwards, while the dashed curves shows the
  orbits integrated forwards. Hermus is shown in blue, while Hyllus is
  shown in red. The Sun's location at (X,Y,Z) =
  (8.5,0,0) kpc is indicated.}

\end{figure}

\end{document}